\documentclass[12pt]{article}      

\usepackage{caption}
\usepackage{fancyhdr}
\usepackage{longtable}
\usepackage{tabu}
\usepackage[section]{placeins}   %
\usepackage[yyyymmdd,hhmmss]{datetime}
\usepackage[pagebackref]{hyperref}  
\hypersetup{ colorlinks,
    citecolor=blue,
    filecolor=blue,
    linkcolor=blue,
    urlcolor=blue
}
\renewcommand*{\backref}[1]{}

\newcommand{\captionl}[1]{\captionsetup{width=.9\textwidth}{\caption[#1]{{\sf #1}}}}
\newcommand{\captionll}[2]{\parbox{15cm}{\caption[#1]{{\sf #2}}}}
\newcommand{\captiontwo}[2]{\parbox{7.5cm}{\caption[#1]{{\sf #2}}}}
\usepackage{xcolor}

\definecolor{gray}{rgb}{0.6,0.6,0.6}
\definecolor{red}{rgb}{0.85,0,0}
\definecolor{green}{rgb}{0,0.85,0}
\definecolor{blue}{rgb}{0,0,0.85}
\definecolor{beige}{rgb}{0.92,0.87,0.78}
\usepackage[all]{hypcap}    

\usepackage{overcite}       
\usepackage{graphicx}
\graphicspath{ {./Figures/} }

\makeatletter \renewcommand\@biblabel[1]{$^{#1}$} \makeatother
 \newlength{\bibhang}
\newcommand{\muenonrhobl}[2]{\left(\overline{\mu_{\rm en}} / \rho
                                                   \right)^{\rm #1}_{\rm #2}}
\newcommand{\muenonrhol}[2]{\left(\mu_{\rm en} / \rho
                                                   \right)^{\rm #1}_{\rm #2}}

\newcommand{\mutronrhol}[2]{\left(\mu_{\rm tr} / \rho
                                                   \right)^{\rm #1}_{\rm #2}}

\newcommand{\cen}[1]{\begin{center} #1 \end{center}}
\newcommand{\eqn}[1]{\begin{equation} #1 \end{equation} }

\usepackage[mathlines]{lineno}

\newcommand{\ie}{{\it i.e.}, }
\newcommand{\eg}{{\it e.g.}, }

\newcommand{\ioc}{{${}^{125}$I}}

\newcommand{\ir}{{${}^{192}$Ir }}
\newcommand{\irc}{{${}^{192}$Ir}}

\newcommand{\ce}{{${}^{137}$Cs }}
\newcommand{\cec}{{${}^{137}$Cs}}
\newcommand{\yb}{{${}^{169}$Yb }}
\newcommand{\ybc}{{${}^{169}$Yb}}
\newcommand{\co}{{${}^{60}$Co }}
\newcommand{\coc}{{${}^{60}$Co}}

\newcommand{\Tv}{{\tt CLRP\_TG43v1}}
\newcommand{\Tvv}{{\tt CLRP\_TG43v2}}

\newcommand{\eb}{{\tt egs\_brachy }}
\newcommand{\ebc}{{\tt egs\_brachy}}
\newcommand{\g}{{\tt g }}

\newcommand{\BD}{{\tt BrachyDose }}

\newcommand{\gofr}{$g(r)$}   
\newcommand{\Fofrt}{$F(r,\theta)$}

\newcommand{\viz}{{\em viz.}}

\newcommand{\Fig}[1]{Fig.~\ref{#1}}

\newcommand{\Figsrange}[2]{Figures~\ref{#1} to~\ref{#2}}

\newcommand{\figh}[5]{   
   \typeout{***************figh********************}
   \begin{figure}[h]
     \begin{center}
     \includegraphics[width=#1cm]{#2}
     \captionll{#4}{#4 \label{#5}}
     \end{center}
   \end{figure}
}
\newcommand{\figtwoh}[8]{ 
   \typeout{***************figtwoh********************}
   \begin{figure}[h]
   \begin{center}
    \begin{minipage}[t]{8.1cm}
     \begin{center}
     \includegraphics[width=8.1cm]{#1}
     \captiontwo{#3}{\label{#4} #3 }
     \end{center}
     \end{minipage}
     \mbox{}\hfill
     \begin{minipage}[t]{8.1cm}
     \begin{center}
     \includegraphics[width=8.1cm]{#5}
     \captiontwo{#7}{\label{#8} #7 }
     \end{center}
     \end{minipage}
   \end{center}
   \end{figure}
}

\newcommand{\figtwoonech}[5]{ 
   \typeout{***************figtwoonech********************}
   \begin{figure}[h]
   \begin{center}
    \begin{minipage}[t]{8.1cm}
     \begin{center}
     \includegraphics[width=8.1cm]{#1}
     \end{center}
     \end{minipage}
     \mbox{}\hfill
     \begin{minipage}[t]{8.1cm}
     \begin{center}
     \includegraphics[width=8.1cm]{#2}
     \end{center}
     \end{minipage}
     \captionll{#4}{#4 \label{#5}}
   \end{center}
   \end{figure}
}

\begin{document}
\cen{{\Large {\bfseries Update  of the CLRP Monte Carlo TG-43  parameter 
database for high-energy brachytherapy sources}} \\
\vspace{5mm}

\pagestyle{empty}
\pagenumbering{roman}
Habib Safigholi$^{a)}$, Marc~J.~P.~Chamberland$^{b)}$, 
Randle~E.~P.~Taylor$^{c)}$,
    Martin~P.~Martinov, D.~W.~O.~Rogers$^{d)}$, and Rowan~M.~Thomson \\
Carleton Laboratory for Radiotherapy Physics (CLRP), Dept. of Physics,\\ 
Carleton University, Ottawa, Ontario, K1S 5B6}
\mbox{~}\vspace{-8mm}\\
$^{a)}$ Present address, Dept of Radiation Oncology, Lady Davis Institute for Medical Research, Jewish General
Hospital, McGill University, Montreal, Quebec, H3T 1E2\\
$^{b)}$ Present address, Medical Physics, The University of
Vermont Medical Center, Burlington, Vermont, 05401\\
$^{c)}$ Present address, Radformation, New York, NY 10017\\
$^{d)}$ Corresponding author, email drogers@physics.carleton.ca\\
\vspace{-2mm}\\
\vspace{-14mm}\\

\begin{abstract}

\mbox{~}\vspace{-11mm}\\

\noindent{\bf Purpose:} To update and extend version 2 of the Carleton
Laboratory for Radiotherapy Physics (CLRP) TG-43 dosimetry database (\Tvv)
for 33 high-energy (HE, $\geq50$~keV)   brachytherapy sources (1 \ybc, 23
\irc, 5 \cec, and 4 \coc) using \ebc, an open-source EGSnrc application.
A comprehensive dataset of TG-43
parameters is compiled, including detailed source descriptions, dose-rate
constants, radial dose functions, 1D and 2D anisotropy functions,
along-away dose-rate tables, Primary and Scatter Separated (PSS) dose
tables, and mean photon energies escaping each source.  The database also
documents the source models which will be freely distributed with \ebc.\\
{\bf Acquisition and Validation Methods:} Datasets are calculated after a
recoding of the source geometries using the egs++ geometry package
and its \eb extensions.  Air kerma per history is calculated in a
10$\times$10$\times$0.05~cm$^3$ voxel located 100~cm from the source along
the transverse axis and then corrected for the lateral and thickness
dimensions of the scoring voxel to give the air kerma on the central axis
at a point 100~cm from the  source's  mid-point. Full-scatter  water
phantoms with varying voxel resolutions in cylindrical coordinates are used
for dose calculations.  Most data (except for \coc) are based on the
assumption of charged particle equilibrium and ignore the potentially large
effects of electron transport very close to the source and dose from
initial beta particles. These effects are evaluated for four representative
sources.
For validation, data are compared to those from \Tv~ and published data.  \\ 
{\bf Data Format and Access:}
Data are available at 
\url{https://physics.carleton.ca/clrp/egs_brachy/seed_database_HDRv2} or 
\url{http://doi.org/10.22215/clrp/tg43v2}
including in Excel (.xlsx) spreadsheets, and  are presented 
graphically in comparisons to previously published data for each source. \\
{\bf Potential Applications:} The \Tvv~ database has applications in 
research, dosimetry, and brachytherapy planning. This comprehensive update
provides the medical physics community with more precise and in some cases
more accurate Monte Carlo (MC) 
TG-43 dose calculation parameters, as well as fully benchmarked and described
source models which are distributed with \ebc. 
\end{abstract}

\vfill
\cen{Table of contents is for drafting and review purposes only.}
\vspace*{-10mm}
\tableofcontents

\vfill
\newpage

\setcounter{page}{1}

\setlength{\baselineskip}{0.7cm}

\pagestyle{fancy}
\pagenumbering{arabic}
\section{Introduction}
\label{intro}

Presently,  brachytherapy dosimetry and planning are often based on the
methodology introduced in the 1995 AAPM Task Group No. 43 report,
(TG-43)~\cite{Na95} and the single source consensus datasets in its various updates and
supplements\cite{Ri04,Ri04a,Ri07a,Ri17,Ri18,Pe12,Pe12a}.  In 2008, the {\bf
C}arleton {\bf L}aboratory for {\bf R}adiotherapy {\bf P}hysics (CLRP)
TG-43 dosimetry database was published for 42 low-energy (LE) and
high-energy (HE)  brachytherapy sources (\Tv)~\cite{TR08b,TR08c,TG43WEB}
using the EGSnrc application \BD\cite{Ta06b}.  These data or source models
have been extensively cited in the
literature~\cite{Ri17,Ri18,Pe12a,TR10,Th08a}.  In 2020, an updated version
of the CLRP TG-43 dosimetry parameters (\Tvv)\cite{Sa20} for 40 LE LDR
brachytherapy sources were calculated utilizing \ebc\cite{Ch16}, an open
source EGSnrc application. This comprehensive \Tvv~database for LE sources
includes dose-rate constants (DRCs), radial dose functions, 1D and 2D
anisotropy functions, along-away dose-rate tables, Primary and Scatter
Separated (PSS) dose tables for some sources, and mean energies of photon
spectra escaping source encapsulation.  The purpose of the present study is
to update, extend, and benchmark a comprehensive TG-43 dosimetry database
for HE brachytherapy sources using a single consistent method using \ebc.
The HE source geometry models were coded using the egs++ geometry
package\cite{Ka20a} and systematically reviewed and improved as needed.  As
of July 2022, twenty new HE sources are added to those in \Tv. Overall, the
updated \Tvv~datasets include 40 LE and 33 HE sources (1 \ybc, 23
\irc, 5 \cec, and 4 \coc) compared to the 2008 \Tv~which includes datasets
for 27 LE and 15 HE sources.

\section{Acquisition and Validation Methods}
\label{methods}

\subsection{Computational tools and Monte Carlo simulations}
\label{methods:mc} 

All Monte Carlo (MC) calculations are performed with the EGSnrc application
\eb \cite{Ch16}  (GitHub commit hash 8166234, 2020, with 
some source models updated since 2020; available at
\url{https://github.com/clrp-code/egs_brachy}).  \eb
is benchmarked and documented in previous publications
\cite{Ch16,Th18,Sa20}. Generally, transport parameters are  EGSnrc defaults
\cite{Ka20}, using the HE default specifications distributed with \ebc.  As
discussed below, electron transport is not generally modelled for \ybc,
\irc, and \ce sources since charged particle
equilibrium~\cite{TR08c,Pe12,Ba09a} is established not far from the sources
and scoring kerma with the tracklength (TL) estimator is much more
efficient\cite{Ch16}. For example, the TL estimator is more efficient by
factors of 200 for \co and 700 for \ir than using energy deposition
scoring.

For \co sources, electrons are tracked and dose is scored using both energy
deposition scoring and the TL kerma estimator. The electron cutoff energy
is set to 10~keV as recommended in the literature~\cite{Pe12,Ba09a}.
Calculations for representative sources [\viz, \yb(4140), \ir (Generic and
mHDR-v2),
\ce(CSM11), and \co (all 4 models)] are done with and without electron
transport (energy deposition versus TL scoring) to establish if there are
differences for DRC, radial dose, and anisotropy functions.  The global
photon energy transport cutoff (PCUT) is set to 1~keV, except for air-kerma
strength calculations for which the cutoff is 10~keV for \ybc,
\irc, \cec, and \co sources as recommended by the AAPM HEBD report~\cite{Pe12}. 
The 10~keV energy cutoff eliminates the contribution of low-energy
characteristic x rays from HE source encapsulations~\cite{Pe12}. 
For three representative sources [\viz, \ybc(4140), LDR \irc(Best), and  HDR
\irc(Generic)] the effect of using a 1~keV cutoff energy on air-kerma
strength  and DRC  values is evaluated.  Reducing PCUT from 10 to 1~keV
does not change significantly ($\leq$0.1\%) the values of $S_K$ and DRC
for the two \ir sources tested.  However, for the \yb source the
corresponding $S_K$ and hence DRC values change significantly (2.8\% and
-2.8\%). The $S_K$ value is increased due to the inclusion of the
low-energy photons which are not attenuated in the vacuum while the dose
in-phantom at the reference point is not affected by these low-energy
photons which are absorbed in the water with a mean-free path of 2~mm or
less. The results presented below (Table~\ref{table:drc_YbCsCo}) are for
the recommended PCUT=10~keV value although the values for PCUT=1~keV are
included in a footnote for the \yb source.

Photoelectric absorption, Rayleigh scattering, Compton
scattering, and fluorescent emission of characteristic x rays are
simulated. Dose is approximated as collision kerma, scored with a
TL estimator in voxels with mass energy absorption coefficients
(distributed with \ebc; calculated with EGSnrc application \g before 2017).
Recent improvements in the \g application\cite{RT19} showed that these
mass energy absorption coefficient values would change by up to a maximum
of 0.2\% using the updated release of EGSnrc. Photon cross sections are
from the XCOM database~\cite{BH87}. The `un-renormalized' photoelectric
cross sections are used as opposed to the renormalized cross
sections\cite{ICRU90}. The energy-fluence-weighted spectrum averaged values
of $\muenonrhobl{}{}$ are proportional to the collision kerma.
Calculations of these $\muenonrhobl{}{}$ values done with either the
normalized or unrenormalized cross sections showed differences of less than
0.02\%, 0.03\%, 0.02\% and 1.0\% for the spectra in-air or in-phantom at
1~cm for \coc, \cec, \irc, and \yb sources respectively.  This difference
is only an issue for the \yb source and even there, the change for the
in-air spectrum vs for the in-phantom spectrum differ by about 0.4\% so
that the effect on the calculated dose-rate constant would be about 0.4\%.
The uncertainty related to cross section uncertainties on other in-phantom
ratio quantities such as \gofr and \Fofrt~ is expected to be even less.

While some issues related to type B uncertainties on the Monte Carlo
results have been addressed, \eg the effects of electron transport, the
effects of the selection of the low-energy cutoff, $\delta$, the effects of
phantom size or uncertainties related to
renormalized vs un-renormalized photoelectric cross sections, the general
issue of overall uncertainties has not been addressed here. These have been
addressed in the past by TG138\cite{tg138}. Their estimates of the overall
$k=1$~ uncertainties on the Monte Carlo calculated values of individual
high-energy TG-43 parameters such as the DRC, \gofr and \Fofrt~ were 2.1\%,
0.5\% and 0.6\% respectively.

Initial photon energies are
sampled from the NNDC~\cite{NNDC} spectra which are distributed with the \eb
package, and also consistent with  the HEBD report~\cite{Pe12,Pe12a}.
Dose calculations are performed with the source located at the center of a
full scattering~\cite{Pe12} cylindrical water phantom (80~cm long, 40~cm
radius, $\rho$ = 0.998~g/cm$^{3}$). Since sources are cylindrically
symmetric, for efficiency purposes, collision kerma and hence absorbed dose
are scored in concentric cylindrical shells. To increase efficiency and also
minimize bin-size artifacts\cite{Ta06b}, the radial and depth resolutions of the
cylindrical shells are 0.1~mm for $r$~$\leq$~1~cm, 0.5~mm for
1~cm~$<$~$r$~$\leq$~5~cm, 1~mm for 5~cm~$<$~$r$~$\leq$~10~cm, and 2~mm for
10~cm~$<$~$r$~$\leq$~20~cm  where $r$ is the radial distance from the
source's central axis. The magnitude of the voxel size effects was
discussed previously \cite{Ta06b,Ba97} and is typically $\leq$
0.25\%\cite{Ta06b}.

Air-kerma per history is always calculated using a
TL estimator for photons above the threshold $\delta$ (10~keV
normally) in a 10$\times$10$\times$0.05~cm$^3$ air voxel
located effectively in vacuo on 
the transverse axis 100~cm away from the source
and then corrected using $k_{r^2} = 1.00217$ for the lateral and thickness
dimensions of the scoring voxel to give the air kerma per history on the
central axis at a point 100~cm from the source's  mid-point. Although the $k_{r^2}$ formula
used here and previously\cite{Ta06b} is
wrong in general, it has been shown to be highly accurate for the thin detector used
here (see ref\cite{Ro19} for details) and so is used here. 
From this, the air-kerma strength per history factor, $S_K^{\rm hist}$, in
Gy$\cdot$cm$^2$/hist, is calculated by multiplying the air kerma on axis
per history by $d^2$, where $d=100$~cm in this case. This factor is useful
when calculating clinical doses and the DRC, $\Lambda$, which is given by
  \eqn{\Lambda = \frac{D(1~\mbox{cm},90^{\circ})/\mbox{hist}}{S_K^{\rm hist}}, ~~~~~~
 \left[\frac{\mbox{Gy/hist}}{\mbox{Gy$\cdot$cm}^2\mbox{/hist}}\right] 
 = \left[\mbox{cm}^{-2} \label{eq_akshf}\right] \equiv 
   \left[\mbox{cGy per hour per U}\right] \label{eq_units}}
where $D$(1~cm,90$^{\circ}$)/hist is the dose to water per history calculated at
the reference point at (1~cm,90$^{\circ}$).
Air kerma is defined in dry air and hence the geometry of the
detector's sensitive region for air-kerma calculations is filled with dry
air as recommended by TG-43U1S2\cite{Ri17}. 

For those phantom voxels which overlapped with the source boundary, dose
is scored only in the portion of the voxel which is not occupied by the source 
and $10^{9}$ or $10^{10}$ random points per cm$^{3}$ are used to determine the
volume correction\cite{Sa20}. The number of photon histories generated in
each simulation was  $\sim 5\times10^{10}$ to ensure that results for DRC, \gofr, and
\Fofrt~(for angular points away from the source's
axis) have  k=1 statistical uncertainties less than 0.02\%,
0.02\%, and 0.1\% respectively. The uncertainty on \Fofrt~ for the 
small voxels very close to the source axis can be up to 4.0\%.

\FloatBarrier

\subsection{TG-43 dosimetry parameters calculations}\label{methods:tg43}

TG-43 dosimetry parameters~\cite{Na95,Ri04} \viz, radial dose function,
$g_{L}(r)$, 1D anisotropy function (anisotropy factor), $\phi_{an}(r)$, 2D
anisotropy function, F(r,$\theta$), air-kerma strength, $S_K$, and dose
rate constant,  DRC or $\Lambda$, for 33 HE brachytherapy sources are
calculated using methods described in our previous work~\cite{Sa20, TR08,
TR08b}. The only significant difference is the inclusion of data to 20~cm
radius compared to 10~cm for the LE database\cite{Sa20}.

\begin{figure}[ht]
\begin{center}
 \includegraphics[scale=0.62]{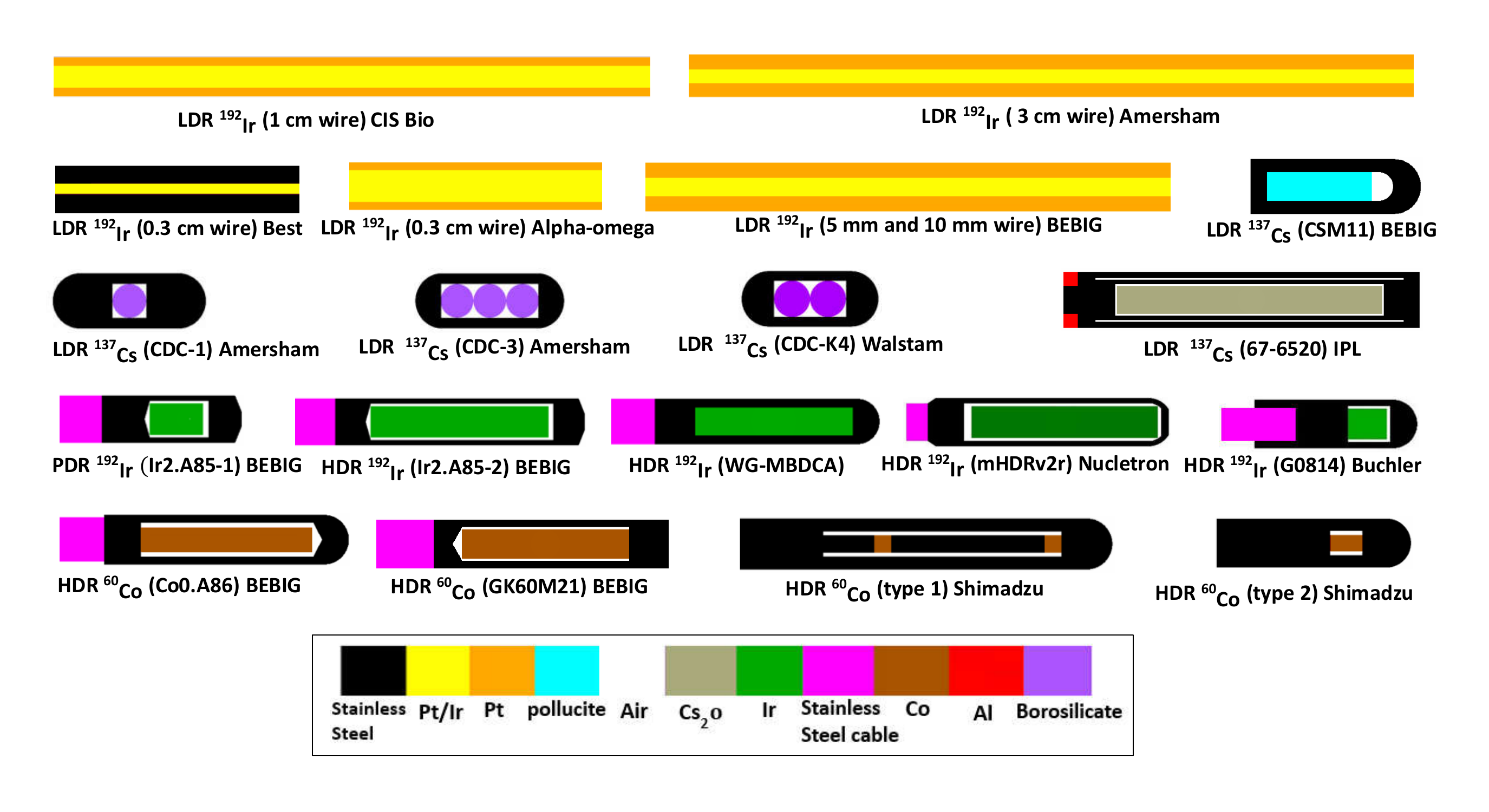}   
\captionl{Schematics of the 20 HE brachytherapy sources added to \Tvv (the
BEBIG \ir LDR source comes in 2 lengths). 
Source geometries
are generated by {\tt egs\_view}, the EGSnrc viewer.  Different
sources have different scales, but each source's dimensions are
individually to scale.
\label{fig_new_sources} }
\end{center}
\end{figure}

The 20 new HE sources
added to the \Tvv~ database are shown in Figure~\ref{fig_new_sources}. A
detailed description of all 33 sources is available online in the
database.
 
\FloatBarrier

\subsection{Effects of electron transport and initial betas}
\label{section-edep-tl}

\subsubsection{Effects on dose-rate constants, $\Lambda$}
\label{section-edep-TL-DRC}
The DRCs calculated with electron transport tend to be larger for the
higher-energy sources since the electrons transport the energy away from
the source in the phantom, thus increasing the dose at the reference point.
At the same time electron transport has no effect on the air-kerma strength
since the air kerma is typically measured  using calibrated ion chambers which are
only sensitive to the photon fluence. Hence the calculated air-kerma
strength per history factor does not involve transport of electrons from
the source.  Specific
examples of differences in dose-rate constants calculated with and without
electron transport in the water phantom are given below in
section~\ref{section-trends} and in the footnotes in
Tables~\ref{table:drc_ir} and ~\ref{table:drc_YbCsCo}.

However there is another issue which appears to be overlooked or ignored in
the literature \viz, the distinction between air collision kerma and air
kerma. Whereas measured values are all of air kerma, as far as we can
determine (many papers lack sufficient information) calculations are all of
air collision kerma based on using $\muenonrhol{}{air}$ values integrated
over the photon energy fluence spectrum in the region of interest. For
low-energy brachy sources this is completely acceptable. For high-energy
sources, using the EGSnrc
application g\cite{Ma20b}, the calculated ratios of
$\mutronrhol{}{air}/\muenonrhol{}{air}$ averaged over the photon energy-fluence
spectra from sources are unity within 0.02\%, 0.04\%, 0.1\%, 0.17\% and
0.33\% for in-vacuum spectra at 100 cm from representative \ioc, \ybc,
\irc, \cec, and \co~sources respectively. In principle, the calculated
air-kerma strength should be increased by these amounts and hence the
dose-rate constants decreased by these amounts. This is only possibly
significant for \ce and \co sources.  Nevertheless, in keeping with what
appears to be all previous practice, these corrections have not been
applied to dose-rate constant values in the database.

\subsubsection{Effects on radial dose function, \gofr }
\label{sec-edep-tl-gofr}

There have been a variety of studies which look at the effects of electron
transport and initial source beta decays on TG-43
parameters\cite{TR08c,Ba09a,BR99,Ba01b,WL02a}.
To show the potential effects of using the tracklength estimator, 
calculations including electron transport have been done for representative 
sources with each radionuclide.
\Fig{fig-gofr-2x2} shows two \gofr~ curves for each of four representative
sources, one calculated with the TL estimator based only on photon
transport and the other using interaction scoring which scores the energy
deposited by electrons along their path. The most dramatic effects occur
within the first few mm (up to 9~mm for \coc). Perhaps the most surprising
result is the very high value very close to the source for the \yb source
which has a mean photon energy of only 110~keV. However, there are some
photons up to 300~keV and the corresponding electrons escaping from the
source deliver considerable dose in the first 0.5~mm. For the
higher-energy sources there are clearly buildup effects in play so that
even the electrons escaping from the source's cladding have not reached
full buildup. Once full buildup is achieved, the electron transport causes a
significant bump in the dose to values greater than those from the TL
estimator at the same radius (7\%,
3\%, and 6\% for the \irc, \cec, and \co sources here respectively). In external
photon beams, the dose does not exhibit a bump relative to the collision kerma. For
these brachytherapy sources the bump is due  to the multiple scattering of
electrons in a radial geometry as demonstrated in \Fig{fig_no_ms} for a
\co point source.  The residual small peak in the
no-scatter case is due to the downstream transport of energy
before being deposited. This affects the ratio for the entire curve but is
maximal at the small radii where the 1/r$^2$ drop off in dose is more
rapid (see further discussion below).
The curve with a high ECUT during the interaction scoring has a value of
1.003 because it is scoring the total kerma rather than the
collision kerma scored by the TL estimator (see
section~\ref{section-edep-TL-DRC}).

Another noteworthy feature of the \gofr~ curves in \Fig{fig-gofr-2x2} is
that the \co TL-scored values past 1~cm are slightly greater than the
interaction-scored values. This is the reverse of the situation in an
external beam\cite{RT19}. However, this is an artifact because of
the normalization at 1~cm where the interaction-scored value is greater
than the TL-scored value.  \Fig{fig_Co_edep_on_tl} demonstrates
this explicitly.

\FloatBarrier
\figtwoonech{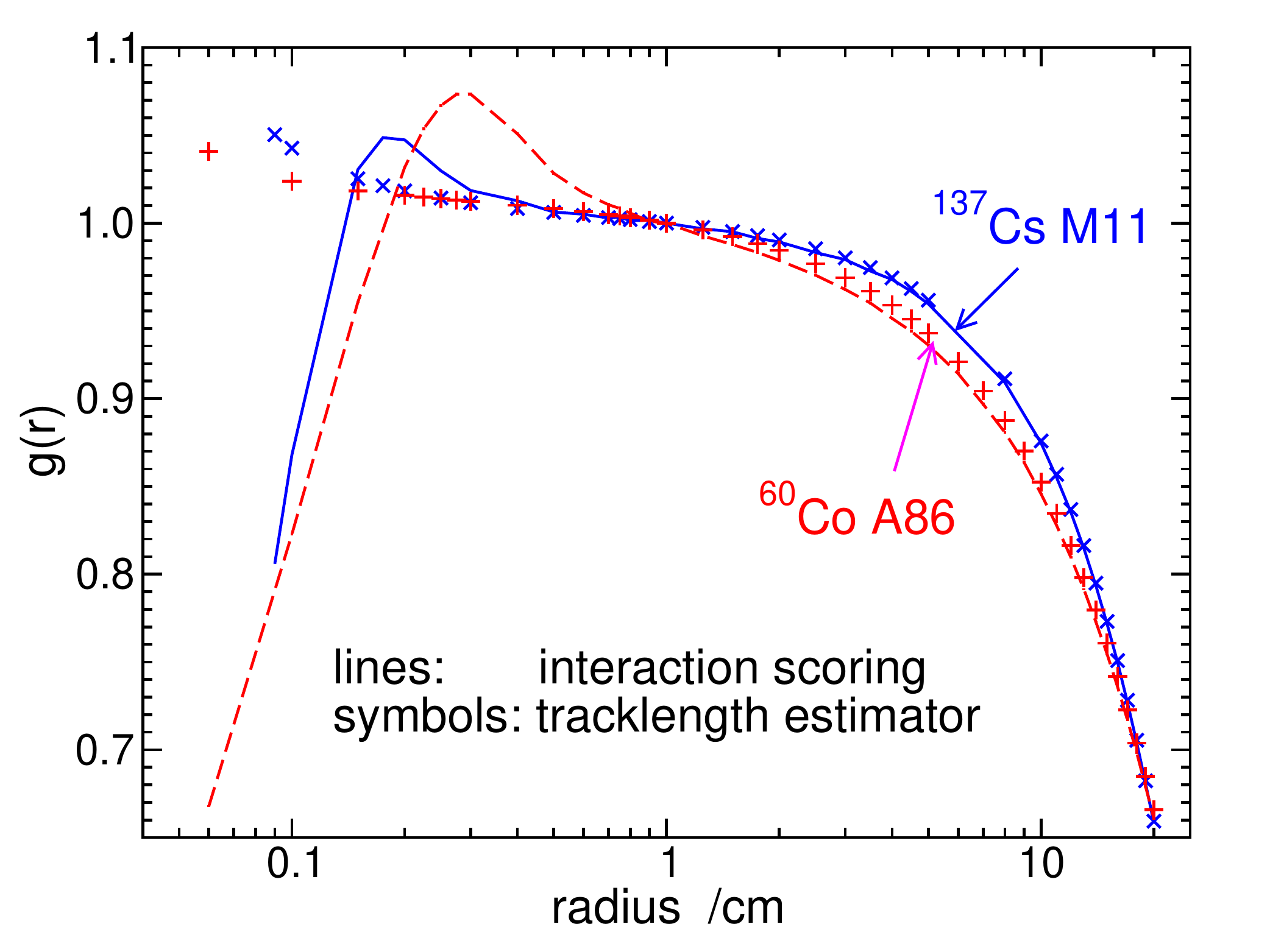}  
   {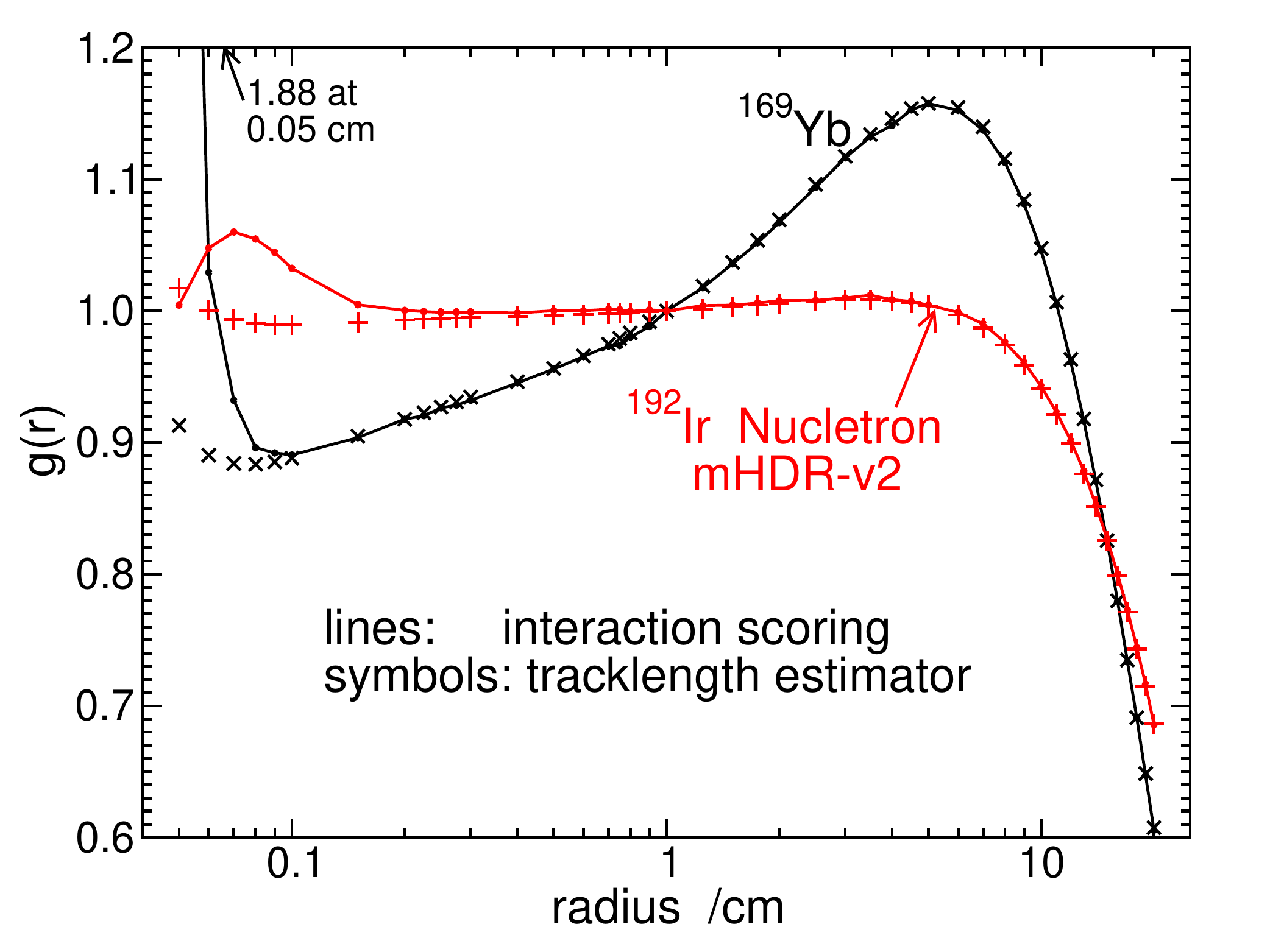}          
   {dummy}
   {\gofr~ values for 4 representative high-energy sources with different
    radionuclides. 
   The line curves (dashed or solid) are calculated scoring energy deposited by
   electrons created by photon interactions and the symbols (x or +) by the
   TL estimator which ignores electron transport effects. No beta
   particles are modelled from the sources.}
   {fig-gofr-2x2}

\FloatBarrier

\figtwoh{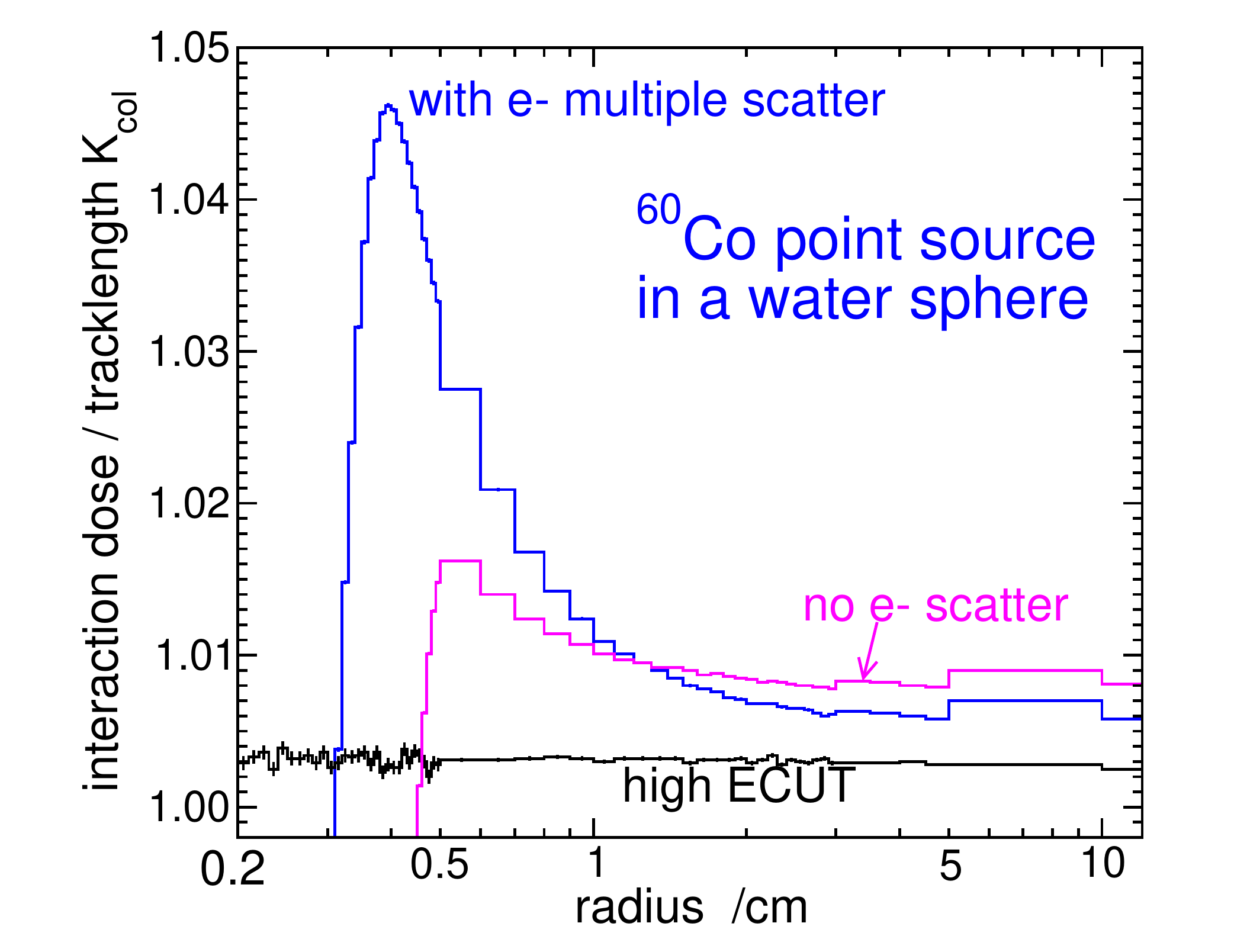}   
   {}
   {Ratio of dose scored with electron transport to the collision kerma
   scored with a TL estimator as a function of radius for a point
   source of 1.25 MeV photons in a sphere of water. The upper curve
   includes the multiple scatter of the electrons set in motion. The curve
   with the lower peak has electron scatter turned off.
   The lowest curve suppressed all electron transport and deposited the
   electron energy where created.}
   {fig_no_ms}
   {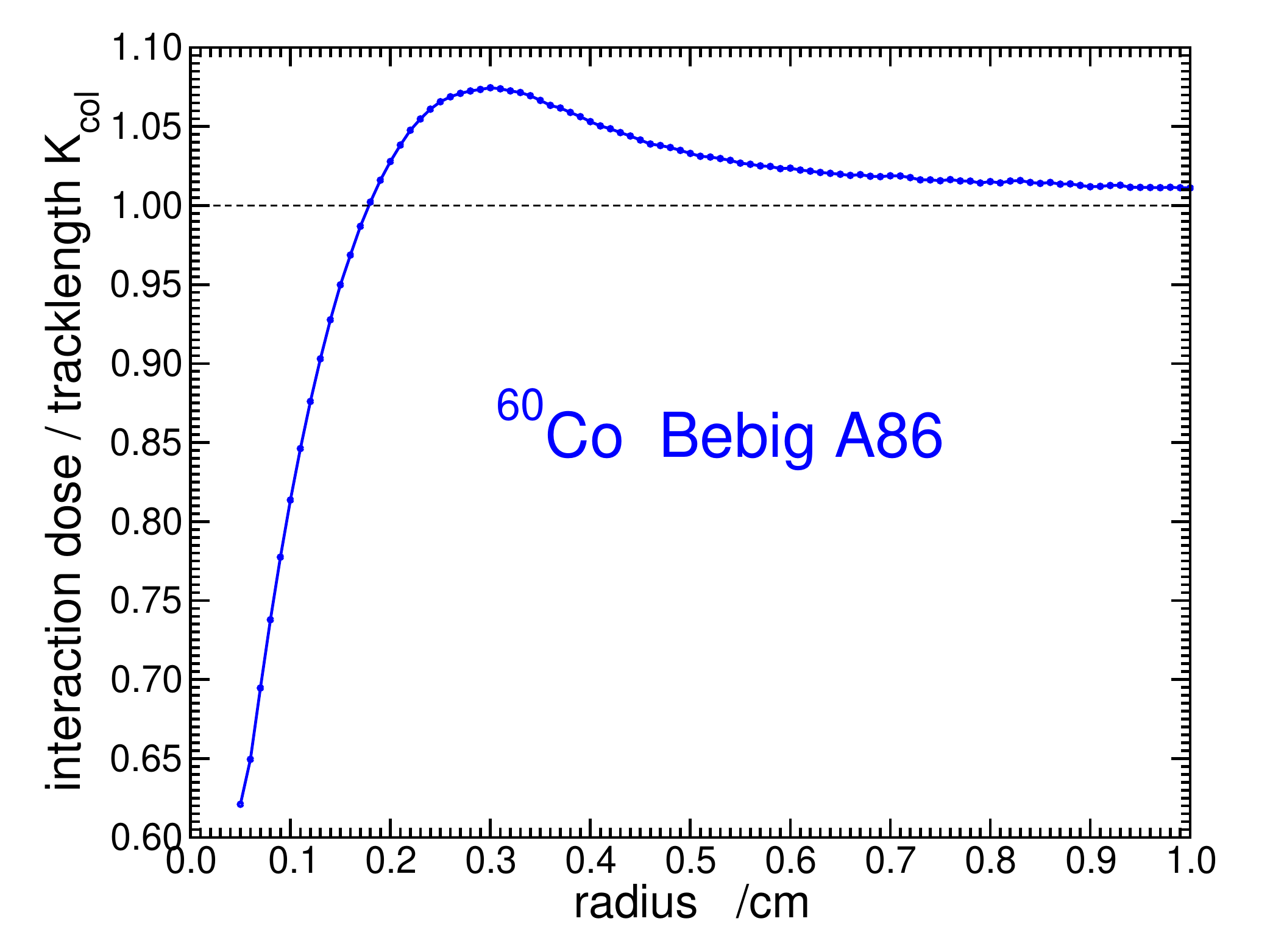}      
  {short}
  {Ratio of the interaction-scored dose to the TL-scored dose
   radially on the transverse axis of the Bebig A86 \co source.}
  {fig_Co_edep_on_tl}

\FloatBarrier
\subsubsection{Electron transport effects on the anisotropy function, 
    \Fofrt }
\label{sec-edep-Fofrt}
For radii close to the source capsule of the Bebig A86 \co source (in this
case, at r=0.26~cm), \Fig{fig-Co60-fofrt} shows that including electron
transport plays a significant role in the anisotropy function, \Fofrt .
The HES curves (High ECUT in Source), which suppress electrons coming from
the source, show that electrons from the source play an important role when
electron transport is accounted for but have virtually no effect when TL
scoring is used. The significant drop in the \Fofrt~ values with electron
transport compared to that evaluated with the TL estimators for low and
high angles is because at these angles the water at radius 0.26~cm is only
0.01~cm from the encapsulation. Hence the electron buildup is not complete
as seen in the \gofr~ curves in \Fig{fig-gofr-2x2} for radii close to the
encapsulation. In contrast, at 90$^\circ$ the dose at a radius of 0.26~cm
is 0.21~cm from the encapsulation and the dose calculated with electron
transport is even higher than that scored with the TL estimator.  Hence the
drop in the electron transport value of \Fofrt~ for $\theta$ away from
90$^\circ$ relative to the TL value is result of the numerator,
$D(r,\theta)$, decreasing  and the denominator, $D(r,90^\circ)$,
increasing.  The right panel of \Fig{fig-Co60-fofrt} shows that even at a
radius of 0.5~cm there is a small but distinct effect from considering
electron transport.

\FloatBarrier
\figtwoonech{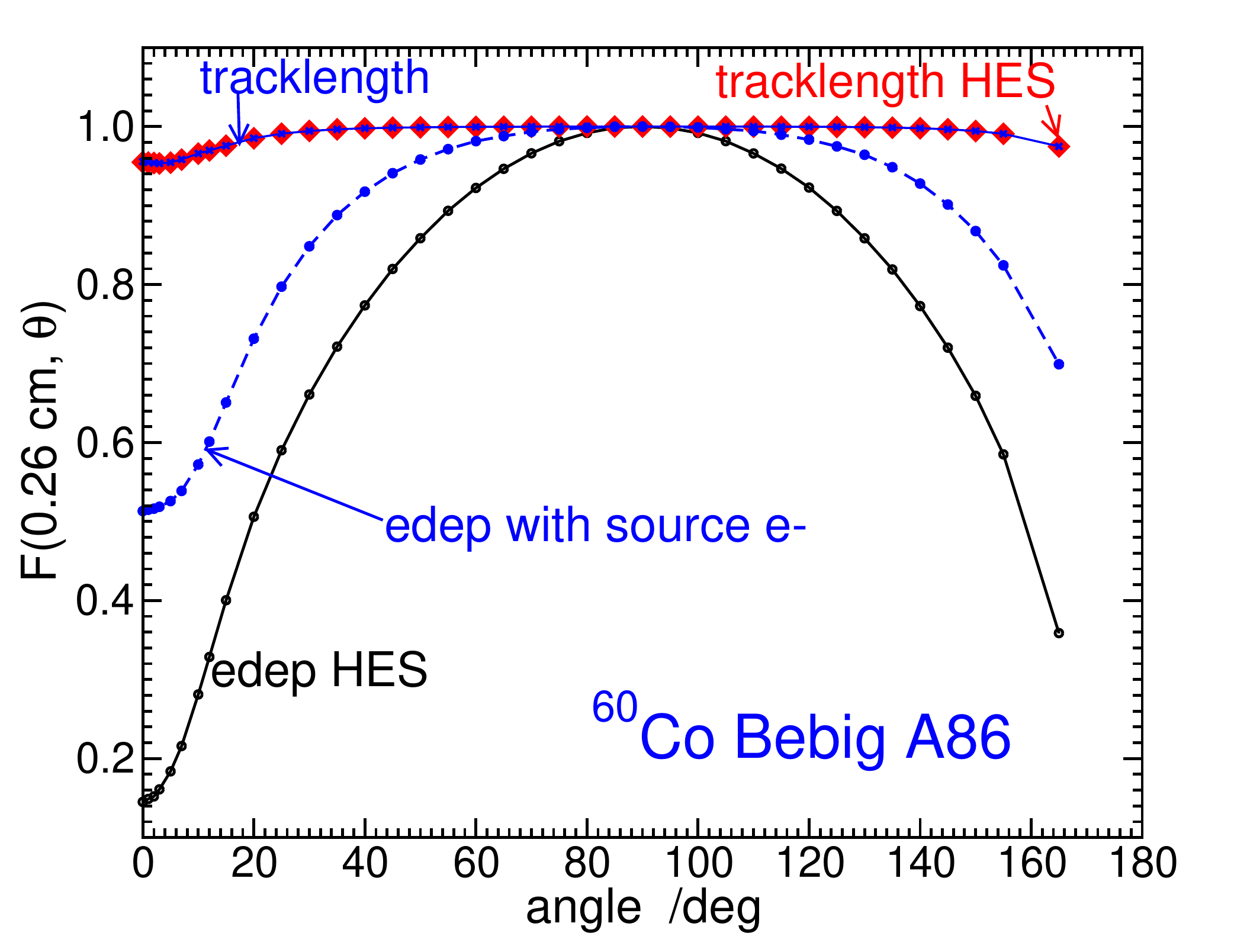}
  {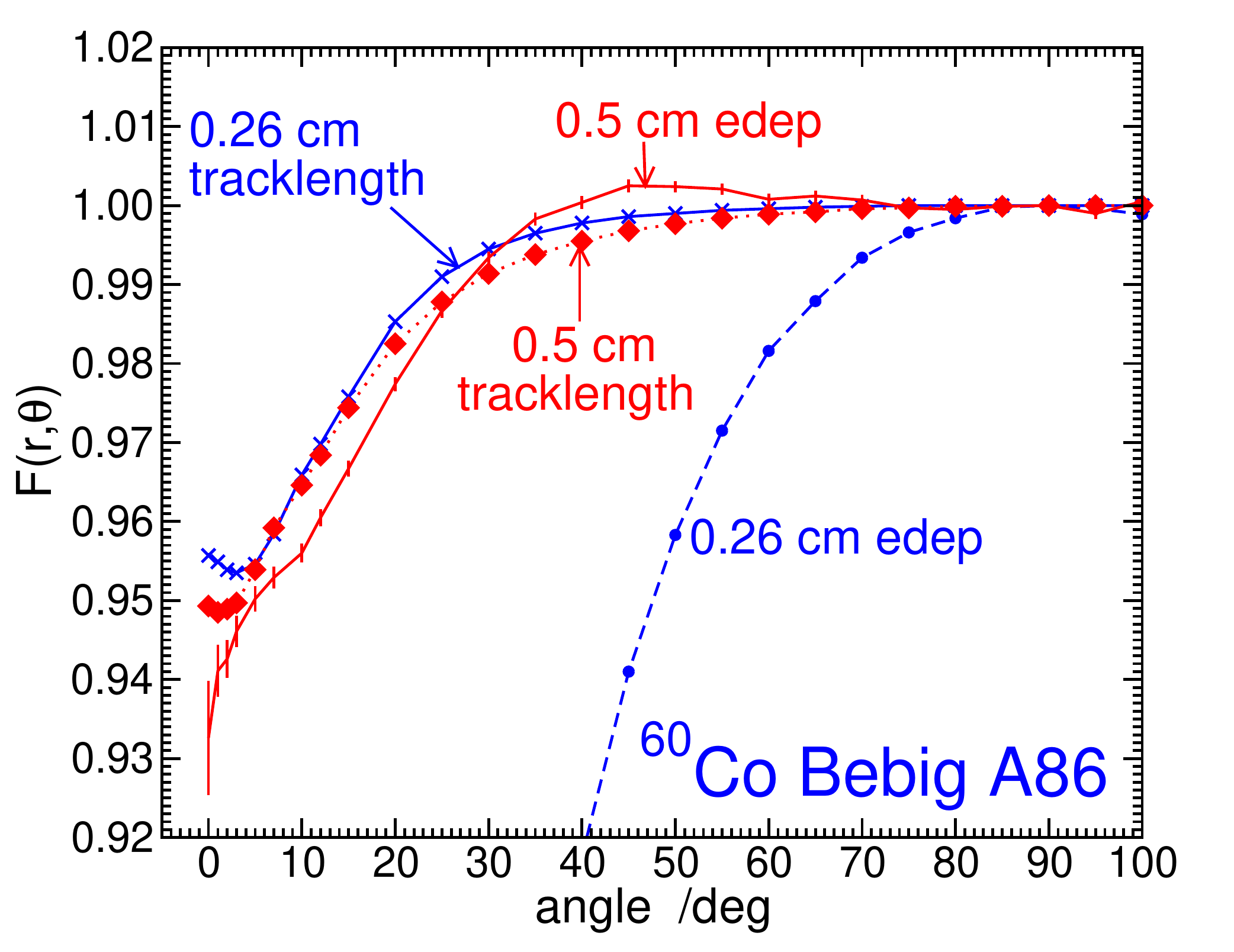}       
  {}
  {Calculated \Fofrt~ values for the Bebig A86 \co source. Left
  panel: all curves are at r=0.26~cm. edep means electrons transported and
  energy deposition scored. HES means ``High ECUT in Source capsule'' so no
  electrons escape the capsule. ``tracklength'' means the TL estimator is
  used and only scores the photons. Right panel shows the detail from 0 to
  100$^\circ$ for radii of 0.26 and 0.5~cm.  The source extends 0.25~cm
  longitudinally and 0.05~cm radially from the mid-point of the active
  layer.}
  {fig-Co60-fofrt}
\FloatBarrier
\Fig{fig-Ir-fofrt} shows that for the MBDCA Generic \ir source\cite{Ba15} the
differences between the two scoring approaches is much less than those for
the \co source although for small radii close to the ends of the source
there are still significant differences which are completely gone by a
radius of 1~cm.

\figh{8}{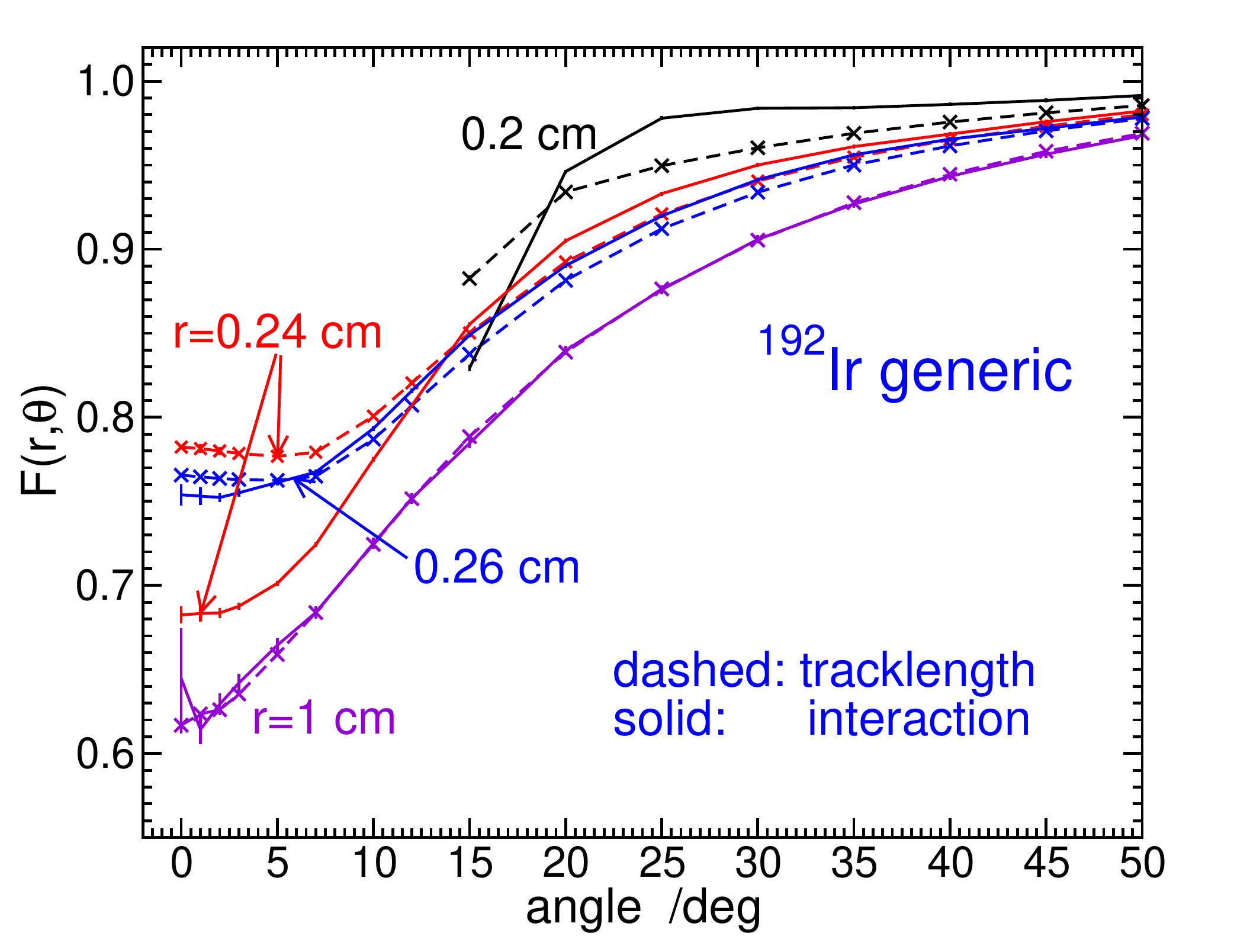}        
   {}
   {Calculated \Fofrt~ values at different radii for the MBDCA
   Generic \ir source using both interaction scoring and the TL
   estimator. This source extends 0.235~cm longitudinally from the mid-point 
   of the radioactivity.}
   {fig-Ir-fofrt}

\FloatBarrier
\Fig{fig-fofrt-Yb} demonstrates that even for the relatively low-energy
\yb source, the effects of electron transport can be significant at locations
near the ends of the source, just as the effect on the \gofr~ curve seen in
\Fig{fig-gofr-2x2} was significant at radii of less than a mm. The high
values for the r=0.24~cm curve close to 0$^\circ$ may reflect slight
problems related to interpolating for values right at the
boundary of the seed.\\
\figh{9}{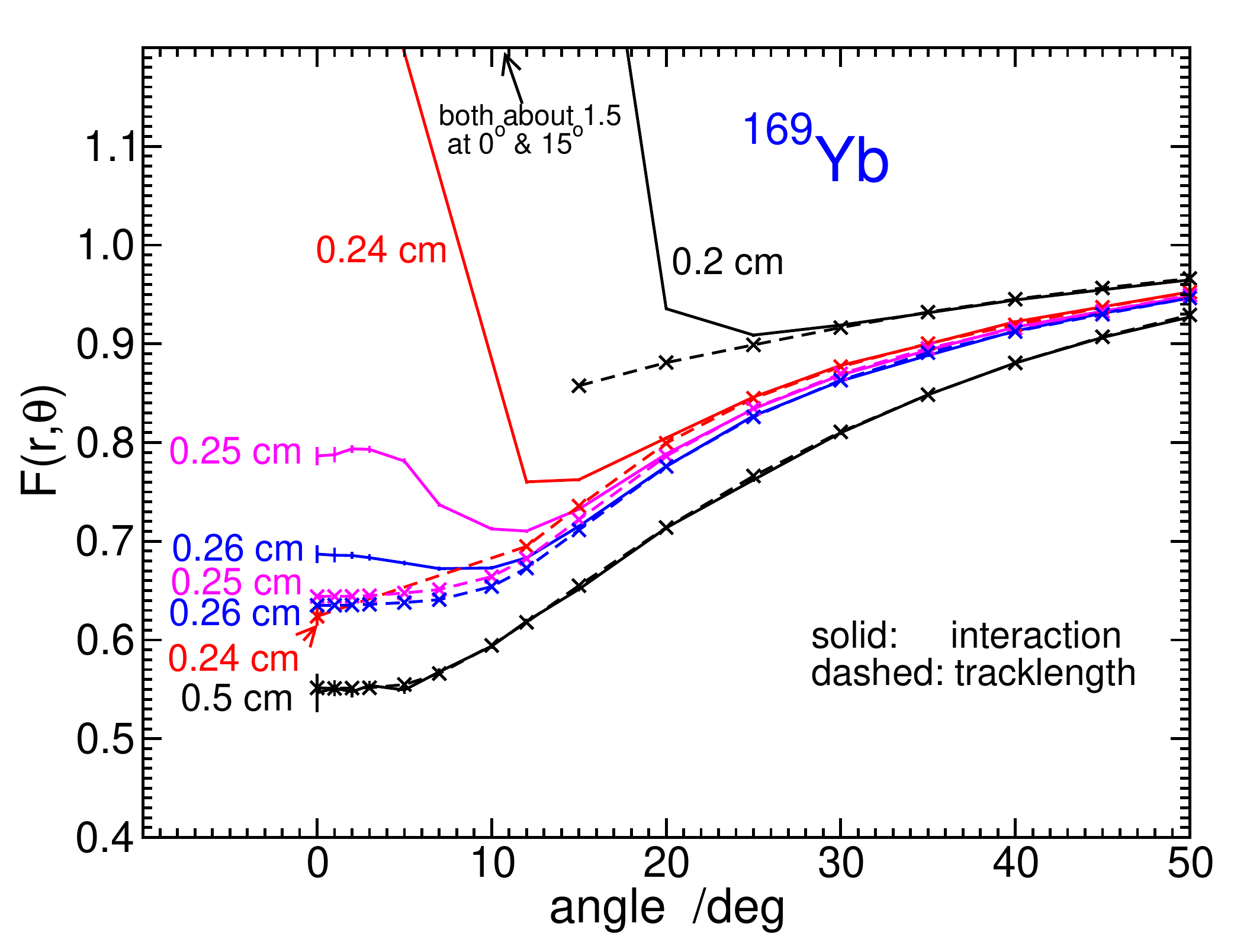}       
  {}
  {Calculated \Fofrt~ values for the 4140 \yb source using
both interaction scoring and the TL estimator. This
source extends longitudinally 0.24~cm from the mid-point of the radioactivity.}
  {fig-fofrt-Yb}
\FloatBarrier

\FloatBarrier
\subsubsection{Effects of initial beta rays and Auger and internal conversion
electrons}
\label{sec:betas}

In addition to the dose from photons,  there are dose contributions from
the betas originating from the radioactive decay 
and from any Auger electrons from the
relaxation of the daughter atom (\eg 0.082 Auger electrons per
disintegration, mostly $<$6~keV for \cec) and any internal conversion
electrons from the relaxation of the daughter nucleus (\eg 0.095 internal
conversion electrons per disintegration with energies from  624 to 661~keV
for \cec)\cite{NNDC}. \Fig{fig-betas} shows the relative dose contribution
from these betas around two \irc, one \cec,  and one \co source. In addition
for the \ce source, it shows the relative dose contribution from the
internal conversion and Auger electrons.  The highest dose contributions
come from the betas that escape the source encapsulation. As seen from the
results for the 2 different \ir sources, the size and extent of this
component depends critically on the geometry of the source. For the very
thin VariSource (r=0.295~mm), the beta dose is much greater than that for
the 69\% thicker MBDCA Generic \ir source(r=0.5~mm) and extends farther
from the VariSource since many more betas escape the source.  Nonetheless,
even for the very thin source this component is negligible by 2~mm from the
central axis. Past that point there is a fairly constant component due to
bremsstrahlung photons created by the betas. For both \ir sources this
constant component is about 0.2\% of the dose from the initial photons and
even less for the \ce and \co sources, consistent with earlier
estimates\cite{BR99,Ba09a}. The peak dose from betas in the \co source is
slightly less than from the other sources but extends much farther radially
although it is less than 0.2\% of the photon dose by a radius of 3~mm. The
surprising aspect of this result is that the increased dose is almost
entirely due to the 0.12\% beta branch which has an endpoint energy of
1491~keV vs the dominant decay with an endpoint energy of 317~keV.

\figh{11}{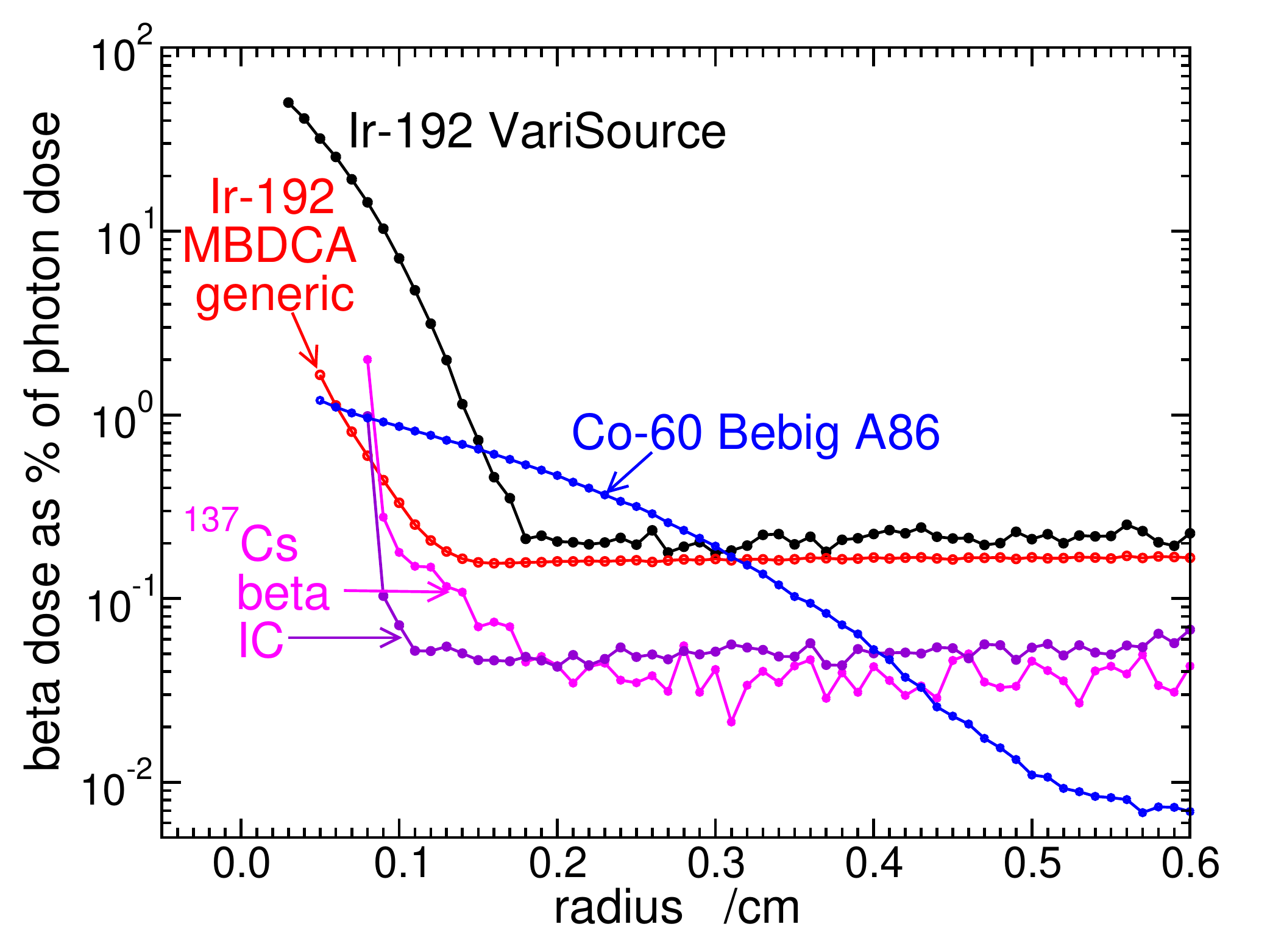}        
  {}
  {The dose on the transverse axis due to initial electrons as a \% of
  photon dose for 4 sources.  The \ir beta spectrum is from ref\cite{DC93}
  as tabulated in ref\cite{BR98}, the \co spectrum from ref\cite{TabRadv3}
  and the other sources from ref\cite{ICRU56}.  There are 2.36 and 2.0
  photons per beta decay in the \ir and \co sources respectively. Internal
  conversion(IC) and Auger electrons are not considered for these 2
  radionuclides. For the \ce source there are 0.929 photons and 0.177
  internal conversion or Auger electrons per beta decay\cite{NNDC}.}
  {fig-betas}
The present results for the effects of betas and other decay products are
generally consistent with the various previous
studies mentioned above\cite{BR99,Ba01b,WL02a,TR08c,Ba09a} given the clear
dependence on the geometries of the sources involved.

\FloatBarrier

\subsubsection{Summary of electron transport effects}
\label{sec-summary-edep}
\Figsrange{fig-gofr-2x2}{fig-betas} in this section~(\ref{section-edep-tl})
show significant dosimetric effects for the four different
radionuclide sources when tracking electrons before scoring energy
deposition and when accounting for the dose due to initial radiations
other than photons.  However, these effects are negligible past a
radius of 2~mm for the \yb and \ir sources or past 3~mm for the \ce
sources. The regions very close to the sources are in the high-dose regions
and the accuracy of the calculations there are not clinically important
(now that intravascular brachytherapy is no longer common). In
the case of \co sources the effect spreads out as far as 6 or 7~mm (see
\Fig{fig_Co_edep_on_tl}) and
potentially are clinically relevant. Thus the calculations for the \co
sources in the database are all reported taking into account electron
transport effects although the 1 or 2\% effects from the initial betas out
to 3~mm radius are not included.

\subsection{Data additional to TG-43 parameters}
\label{methods:additional data}

For all HE sources, along and away dose-rate data normalized to the
air-kerma strength, $S_K$, are tabulated in cGy h$^{-1}$ U$^{-1}$ at 16
away distances from 0 to 20~cm and 29 along points from -20 to 20~cm.  The
away data are only needed for values greater than zero due to the
cylindrical symmetry of the sources.  Primary and Scatter Separated
(PSS)~\cite{TR08c} dose data  are provided.  These data include total,
primary, single scatter and multiple scatter in polar coordinates. They are
normalized to the total photon energy escaping from the source capsule and
tabulated at 12 radii from 0.30 to 20~cm and 47 polar angles with
resolution of $5^{\circ}$ or better. High resolution ($\Delta$r = 1~mm,
$\Delta\theta=1^{\circ}$) PSS data are also provided.  For the purpose of
these calculations, any photon escaping the source encapsulation is
considered a primary photon, only photons which scatter within the phantom
are counted in the scatter tallies.  In Tables~\ref{table:drc_ir} and
\ref{table:drc_YbCsCo} the mean energy of photons escaping the source
surface are provided in column 1 (in brackets).  Calculated photon fluences
as a function of energy have statistical uncertainties less than 0.2\%.

\FloatBarrier
\newpage
\subsection{Data validation}\label{methods:validation}

\begin{figure}[ht]
   \begin{center}
\includegraphics[scale=0.6]{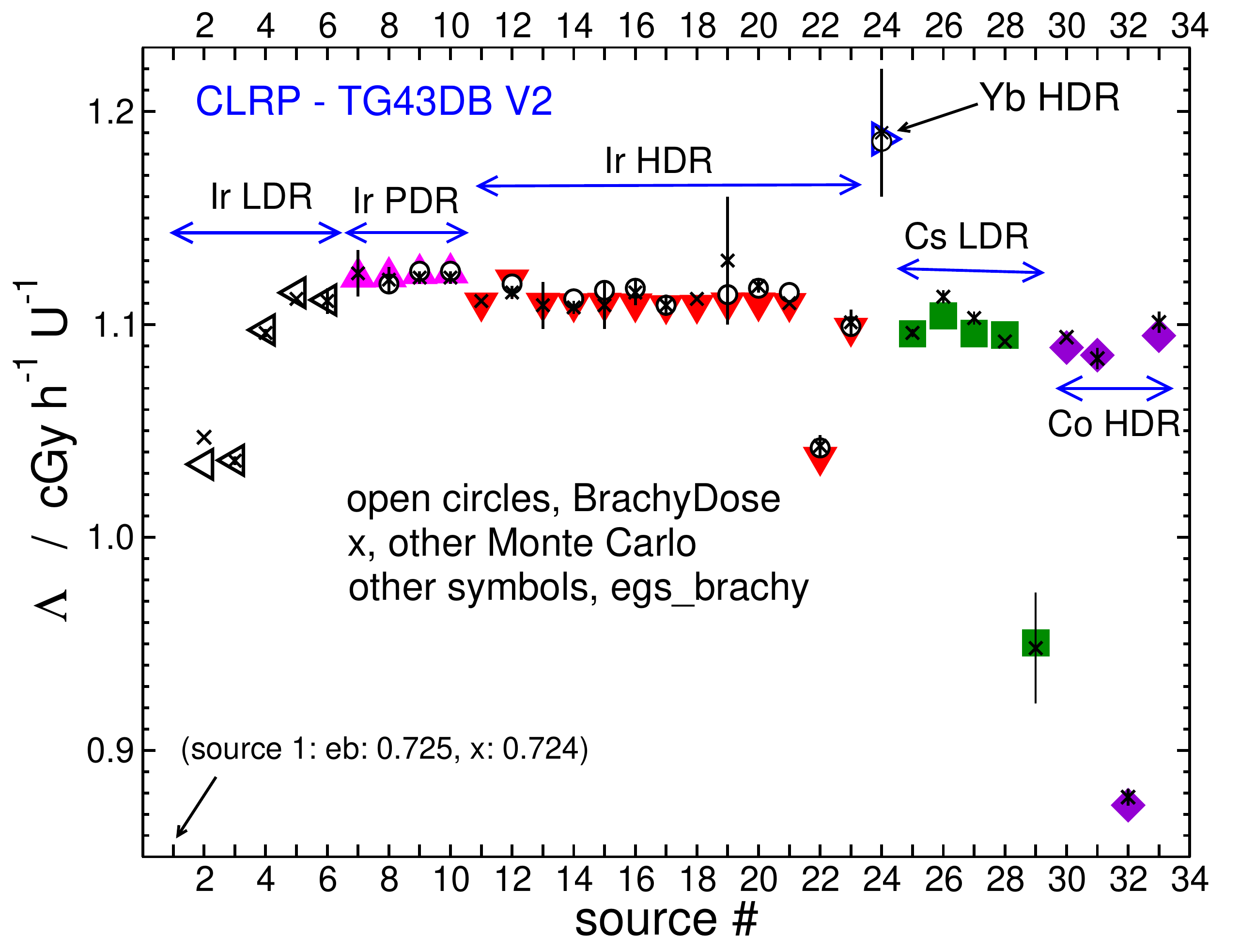}     

\captionl{\eb~calculated values of dose-rate constants, $\Lambda$, are
shown as: black left triangles for \ir LDR, magenta up triangles 
for \ir PDR, red down triangles for \ir HDR,  a blue right triangle for
$^{169}$Yb HDR, green squares for $^{137}$Cs LDR, and  violet diamonds for
\co HDR sources.  Values calculated previously with \BD  are open black
circles~\cite{TR08c,TR08bweb}. DRCs from other MC codes with their
uncertainties are shown as black x symbols. The statistical uncertainties on
\eb and \BD values are smaller than their symbols. The x-axis source
numbers are in Tables ~\ref{table:drc_ir} and \ref{table:drc_YbCsCo}.
\label{fig_DRCcompare} }
  \end{center}
  \end{figure}

For validation, TG-43 dosimetry parameters are compared to 
data in the literature. DRC values and their statistical
uncertainties  are compared to the values computed  with \BD in the
\Tv~\cite{TR08b,TR08c,TG43WEB} database, as well as MC data by other
authors cited in Tables~\ref{table:drc_ir} and~\ref{table:drc_YbCsCo}.
It should be emphasized that although \eb and \BD are based on EGSnrc, the
coding of the geometries is based on completely different geometry packages
and hence the comparison is a validation of the geometry models.
The comparisons give percent differences as:
  \begin{equation}
   \%\Delta(\Lambda_1,\Lambda_2) = \frac{\Lambda_1-\Lambda_2}{\Lambda_1} 
   \times 100\%.
   \label{eq_Delta}
  \end{equation}
DRC values separated by radionuclide are shown in
 Table~\ref{table:drc_ir} and
\ref{table:drc_YbCsCo} and Figure~\ref{fig_DRCcompare}.  Due to the large amount of data generated in this work,
detailed comparisons of \gofr~ and \Fofrt~ are omitted here, but the
comparisons are available in the database whenever other data are
available.

Overall, the CLRPv2 DRC values show good agreement with the available \BD
data with a maximum difference of 0.6\% from the
\Tv~\cite{TR08b,TR08c,TG43WEB} database.  The average difference between
\Tv~ and `v2' for twelve sources is $(-0.19\pm0.24)$\% which is excellent
agreement considering the statistical uncertainty of 0.3\% on the \BD
results~\cite{TR08b,TR08c,TG43WEB}.
This excludes the DRC values for two HDR \ir sources (Flexisource and
MicroSelectron-v2) where some small geometry changes were found necessary
and explained in each source's webpage.  The  resulting changes in the DRC
values (-0.53\%, and 0.0\%) were less than the  maximum 0.6\% difference
between the two databases.

\begin{table} [h]
\begin{center}
\caption{\sf Dose-rate constant values for \ir sources, calculated by \eb 
(`This work'), \BD (`TR')\cite{TR08c, TR08bweb}, other
codes (`Other MC'), and TG-229's TG-43 consensus data\cite{Pe12a} (TG43$_{\rm
con}$). All \eb and \BD values are without electron transport. Statistical
uncertainties are $\le$0.3\% (BD), $\le$0.02\% (eb), and otherwise shown in
brackets (uncertainty in last digit). Percent differences are given
between results for \eb and the best \BD geometry [$\%\Delta$(eb,BD)], as
well as with TG43$_{\rm con}$ { [$\%\Delta$ (eb,TG43$_{\rm
con}$)]}.  The mean emitted photon energy ($\bar{E}_\gamma$) on the surface of
each source determined by \eb is indicated. Sources are numbered for
reference in later figures. Webpages often report several other MC results.
\vspace{-5mm}\\
\label{table:drc_ir}}
\begin{tabular}{lcccccc}
\hline
& \multicolumn{6}{c}{Dose-rate constant, DRC or $\Lambda$ (cGy h$^{-1}$ U$^{-1}$)}\\
\cline{2-7}
\vspace{-0.5mm}{Source model} ($\bar{E}_\gamma$/keV)& This  &
TR~\cite{TR08c, TR08bweb} & {$\%\Delta$} & {Other} &  \footnotesize{TG43$_{\rm con}$}\cite{Pe12a}  & {$\%\Delta$}\\
&  work(eb) &  {BD} & \footnotesize{(eb,BD)} &
{MC} &  & \footnotesize(eb,TG43$_{\rm con}$)\\
\hline
 {\bf $^{192}$Ir LDR}  	    &	   &	   	&    	&		&      &	\\
1  Amersham, 3~cm wire(358.4) 	&	0.7245	&	-	& 	-	&	 0.724~\cite{Ka01c}  & - & - \\

2 CIS Bio, 1~cm wire(360.9)	&	1.0343 &	-	& 	-	&	 1.047~\cite{Pe99}  & - & - \\
~
3 BEBIG, 1~cm wire(358.1)	  &	   1.0362 &	-	& 	-	&	 1.036(2)~\cite{La08}  & 1.036(2) & 0.0 \\

4 BEBIG, 0.5~cm wire(358.2)	  &	   1.0974 &	-	& 	-	&	 1.096(2)~\cite{La08}  & 1.096(2) & 0.1 \\

5 Best, 0.3~cm wire(355.4)	  &	   1.1149 &	-	& 	-	&	 1.112(1)~\cite{Ba04b}  & 1.110(15) & 0.4 \\

6 Alpha-Omega, 0.3~cm wire(360.7)&	 1.1115 &	-	& 	-	&	1.111(1)~\cite{Ba04b} &	- 	& 	- \vspace{2mm}\\

{\bf $^{192}$Ir PDR} 	   &	  &  	&	&	&  &\\

7 BEBIG, Ir2.A85-1(361.2)  &	 1.1222 &	-	& 	-
&1.124(11)~\cite{Gr08a} &	1.124(1) 	& 	-0.2 \\

8 Nucletron, mPDR-v1(359.3)  &	 1.1223 &	1.119	& 	0.3	&	1.121(6)~\cite{Ka03b} &	1.120(6) 	& 	0.2 \\

9 Varian, GammaMed 12i(359.0)&	 1.1239 & 1.125	& -0.1	&
1.122(3)~\cite{Pe01b} &	1.126(3) 	& 	-0.2 \\

10 Varian, GammaMed Plus(359.0)  &	 1.1242 & 1.125	& -0.1	& 1.122(3)~\cite{Pe01b} &	1.123(3) 	& 	0.1 \vspace{2mm}\\

{\bf $^{192}$Ir HDR} 	   &	  &  	&	&	&  &\\

11 Generic, WG-MBDCA(360.6)  &	 1.1102$^a$ &  -	&  -	& 1.111(4)~\cite{Ba15}$^b$ &  -	&  - \\

12 Buchler, G0814(362.3)  &	 1.1211 &  1.119	&  0.2	& 1.115(3)~\cite{Ba01}  & 	1.117(4)	&  0.4 \\

13 BEBIG, Ir2.A85-2(360.5)  &	 1.1102 &  -	&  -	& 1.109(11)~\cite{Gr08a} & 	1.109(12)	&  0.1 \\

14 BEBIG, GI192M11(360.5)  &	 1.1099 &  1.112	&  -0.2	& 1.108(3)~\cite{Gr05b} & 	1.110(4)	&  0.0 \\

15 Nucletron, Flexisource(360.4)  &	 1.1101$^c$ &  1.116	&  -0.5	&
1.109(11)~\cite{Gr06} & 	1.113(11)	&  -0.3 \\

16 Nucletron, mHDR-v1(360.6)  &	 1.1099 &  1.117	&  -0.6	&
1.115(6)~\cite{WL95} & 	1.116(9)	&  -0.5 \\

17 Nucletron, mHDR-v2(361.1)  &	 1.1090$^{c,e}$ &  1.109   &  0.0	&
                  1.107(8)~\cite{Gr11}$^{,d}$ & 	1.109(12)	&  0.2 \\

18 Nucletron, mHDR-v2r(360.5)$^{f}$  &	 1.1092 &  -	&  -	&
                           1.1121(8)~\cite{Gr11}$^{,d}$ & 	-	&  - \\

19 Oncology system, M19(361.0)  &	 1.1103 &  1.114	&  -0.3	& 1.130(3)~\cite{MM07} & 	- 	&  - \\

20 Varian, GammaMed 12i(360.5)  &	 1.1103 &  1.117	&  -0.6	&
1.118(3)~\cite{Ba01} & 	1.118(3) 	&  -0.7 \\

21 Varian, GammaMed Plus(360.5)  &	 1.1100 &  1.115	&  -0.4	&
1.110(1)~\cite{Wu21} & 	1.117(5) 	&  -0.6 \\

22 Varian, VariSource(357.7)  &	 1.0378 &  1.042	&  -0.4	& 1.043(5)~\cite{Ka00c} & 	- 	&  - \\

23 Varian, VS2000(357.7)  &	 1.0984 &  1.099	&  -0.1	&
1.101(6)~\cite{An00d} & 	1.100(6) 	&  -0.1 \\

\hline

\end{tabular}
\end{center}
\mbox{~}\vspace{-5mm}\\
$^a$ \scriptsize{Value calculated using TL estimator. Value 
 including electron transport is 1.1115(17), \ie ($0.12\pm0.15$)\% higher}\\
$^b$ \scriptsize{Value is averaged over 5 different Monte Carlo codes~\cite{Ba15}}\\
$^c$ \scriptsize{There is a change in the model of the source described in the database}\\
$^d$ \scriptsize{Value is averaged over 3 different Monte Carlo codes~\cite{Gr11}}\\
$^e$ \scriptsize{Value calculated using TL estimator. Value including 
electron transport is 1.1077(12), \ie ($0.12\pm0.11$)\% lower.}\\
$^f$ \scriptsize{Elekta sells this source as  mHDR-v2 and no longer sells
the original mHDR-v2.}
 
\end{table}

\clearpage

\begin{table}


\typeout{***********start table 2 ********************}
\begin{center}
\caption{\sf Same as Table~\ref{table:drc_ir},
except for \yb HDR, \ce LDR, and \co HDR sources.  
\label{table:drc_YbCsCo}}
\begin{tabular}{lccccccc}
\hline
& \multicolumn{6}{c}{Dose-rate constant, DRC or $\Lambda$ (cGy h$^{-1}$
U$^{-1}$)} \\
\cline{2-6}
\vspace{-0.5mm}{Source model} ($\bar{E}_\gamma$/keV)& This  &  
Other &  $\%\Delta$  &
TG43$_{\rm con}$\cite{Pe12a}  & {$\%\Delta$}\\
&  work(eb) &  
MC & {(eb,Other)}  &  & (eb,TG43$_{\rm con}$)\\
\hline
 {\bf $^{169}$Yb HDR}     &   &	&    	&		&      &	\\
24  Implant Sciences, 4140 (117.07)$^a$ 	&	1.1871$^{b,c}$ 	&
1.19(3)~\cite{Me06a} &- & - & - \vspace{2mm}\\

 {\bf $^{137}$Cs LDR}  	    &	   &	   	&    	      &	\\
 
25 BEBIG, CSM11(647.64)	  &	   1.0956$^d$ &
1.096(2)~\cite{Ba00a}  & -0.03 & 1.094(18) & 0.1 \\

26 Amersham, CDC-1(646.34)	  &	   1.1040 &
1.113(3)~\cite{Pe02a}  & -0.8 & - & - \\

27 Amersham, CDC-3(647.46)	  &	   1.0959 &
1.103(3)~\cite{Pe02a}  & -0.6 & - & - \\

28 Walstam, CDC-k4(643.74)	  &	   1.0949 &		 1.092(1)~\cite{Pe01d}  & - & - \\

29 IPL,67-6520(646.3)  &	 0.9505 &
0.948(26)~\cite{Me09a}& 0.3  &	0.948(3) & 	0.3 \vspace{2mm}\\

{\bf $^{60}$Co HDR} 	   &	  &  		&  &\\

30 BEBIG, Co0.A86 (1240.48)  &	 1.0985$^e$ &	 1.094(3)~\cite{SB10} 
& 0.4 &	1.092(5) 	& 	0.6 \\

31 BEBIG, GK60M21 (1239)  &	 1.0966$^e$ &	 1.093(2)~\cite{SB10} 
&  0.3 &	1.089(5) 	& 	0.7 \\

32 Shimadzu, Ralston type 1(1210.4) &	 0.8822$^e$ &	 0.878(4)~\cite{Pa03c}
& 0.5  &	- 	& 	- \\

33 Shimadzu, Ralston type 2 (1211.22) &	 1.1077$^e$ &	 1.101(5)~\cite{Pa03c}
& 0.6  &	- 	& 	-\\

\hline

\end{tabular}
\end{center}
\mbox{~}\vspace{-5mm}\\
$^a$ \scriptsize{The BD\cite{TR08c, TR08bweb} and $\%\Delta$(eb, BD) values
are 1.186 and 0.09\%, respectively}\\
$^b$ \scriptsize{Value calculated using TL estimator. Value 
including electron transport is 1.1874(8), \ie 0.03\% larger}\\
$^c$ \scriptsize{Value is for PCUT = 10 keV. Value for PCUT = 1 keV is
1.1541(4), \ie 2.8\% lower} \\
$^d$ \scriptsize{Value calculated using TL estimator. Value 
including electron transport is 1.0990(12), \ie 0.31\% larger}\\ 
$^e$ \scriptsize{Values calculated using electron transport. Corresponding values using
TL estimator are 1.0845(2) (Co.A86, 1.3\% lower), 1.0855(2) (GK60M21, 1.0\%
lower), 0.8742(2) (Ralston type 1,0.9\% lower), and 1.0947(2) (Ralston type
2,1.2\% lower), respectively} \\

\end{table}

\clearpage

\FloatBarrier
   
\FloatBarrier

TG-229~\cite{Pe12a} recommended consensus DRC values
for HE sources based on MC values. The
last column of Tables~\ref{table:drc_ir} and~\ref{table:drc_YbCsCo}
indicate good agreement between the \eb results and the consensus values.
The mean difference between \Tvv~ and
TG-43$_{\rm con}$ DRC values for different source types are:  
LDR, 0.2\%; PDR and HDR \irc, -0.01\%;  LDR \cec, -0.02\%; and HDR
\coc, 0.6\%:  \ie agreement is excellent where comparisons are available. 
Overall, $g_L(r)$  and F$(r,\theta)$ values generated from the \eb data
are in good agreement with the \BD results within statistical uncertainties
for all sources studied by both. As discussed previously\cite{Sa20}, significant
discrepancies in the calculated doses occurred for some regions very close
to the source since \BD did not properly account for statistical
uncertainties in the source volume correction whereas these are now properly
handled in \ebc.

\subsection{Trends seen in the data}
\label{section-trends}
In general, for \ir sources, as the source length increases above the
common 3 to 4~mm range, the DRC values decrease (LDR sources 1-4, HDR
sources 22, 23) with the smallest DRC of 0.7245 being for the 3~cm long LDR \ir
Amersham source (source 1, Table~\ref{table:drc_ir} and
Figure~\ref{fig_DRCcompare}). This is due to the increase in the mean
photon pathlength from the source to the reference point at (1~cm,
90$^{\circ}$). This leads to increased attenuation in the water phantom but
not in vacuum for the $S_K$ geometry and hence the lower DRC.  The same 
applies for the relatively
long \ce IPL source (source 29, Table~\ref{table:drc_YbCsCo} and
Figure~\ref{fig_DRCcompare}) which has the smallest DRC of the \ce sources.

\begin{figure}[ht]
\begin{center}
\includegraphics[scale=0.44]{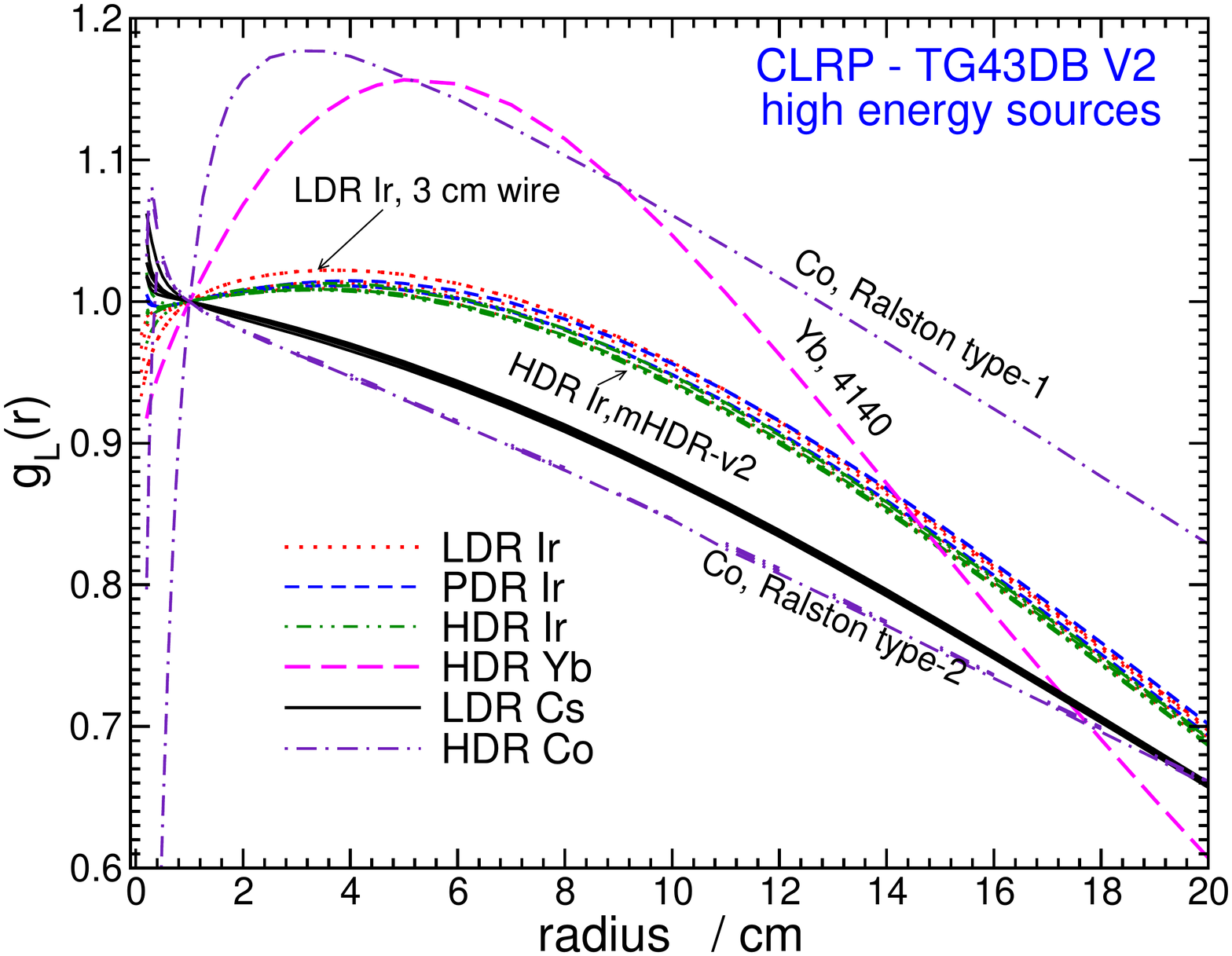}

\captionl{Radial dose functions, $g_L(r)$, of all 33 HE sources including 6
LDR \ir (orange dotted lines), 4 PDR \ir (blue dashed lines), and 13 HDR
\ir (green dash-dotted lines), 1 HDR \yb (pink dashed lines), 5 LDR \cec
(nearly identical solid black lines making very thick black line), 
and 4 HDR \co (violet dash-dotted lines). Only \co
sources used electron transport.
\label{fig_gofr}}
\end{center} 
\end{figure}

The highest DRC among all the  HE and LE sources belongs to the \yb source
with the lowest mean energy of the HE sources (117.1~keV). 
At these energies there is far more scatter at 1~cm than for higher
energy sources. At the reference position, (1~cm, 90$^{\circ}$), 33\% 
of the dose is from
scattered photons whereas for the source with the next highest DRC (source
12) only 10.5\% of the dose at (1~cm, 90$^{\circ}$) is from scattered
photons.  The large amount of scattered dose also leads to \ybc's very large
\gofr~ value for $r$ values near 5~cm (see discussion below re 
Fig~\ref{fig_gofr}).

Among the \co sources the lowest DRC value is for the Shimadzu Ralston type
1 model (source 32). This is due to the source geometry
(Figure~\ref{fig_new_sources}) in which two symmetric active pellets are
spaced by  9~mm of stainless steel which makes it effectively a long source
with a low DRC. The lack of radioactivity near the centre of the source
leads to an unusual \gofr~ which reaches a high value near r=3~cm but this
is due to the low relative dose rate at 1~cm and not excessive scatter as
observed with the \yb source.

For \co sources and some other representative HE sources, DRC values are
calculated with electron transport included for the in-phantom calculations
rather than using the TL estimator.  As mentioned in
section~\ref{section-edep-TL-DRC}, the DRCs calculated with electron
transport are larger than those with the TL estimator.
Differences in DRCs with and without electron transport were $\leq$ 0.2\%,
0.2\%, 0.12\% and 0.3\% for the \yb (4140), \ir (Generic), \ir (mHDR-v2) and \ce (CSM11)
sources, respectively (see footnotes in Tables~\ref{table:drc_ir} and
~\ref{table:drc_YbCsCo}). For all \co source models, the reported DRC
includes electron transport. The mean of these DRC values is 1.1\% higher
than the mean based on the TL estimator (see footnote `e' in
Table~\ref{table:drc_YbCsCo}).

Figure~\ref{fig_gofr} provides $g_L(r)$ values as a function of $r$ for all
33 HE sources(see section~\ref{sec-edep-tl-gofr} for a discussion of
electron transport effects for small radii).  For sources of the same radionuclide, $g_L(r>$1~cm) values
are very similar except for  differences between the Ralston type 1 or type
2 \co sources for which the two radioactive pellets are either placed at
the source center (class B source) or spaced by a 9~mm Stainless steel rod
(class A source)\cite{TR08}. The $g_L(r)$ values for the \yb source are
greater than those for other radionuclides. As discussed above regarding its
large DRC (section~\ref{section-trends}), this is because of the large dose
from scattered photons which are back and side-scattered for photon
energies near 100~keV more than at higher energies. This is seen clearly in
the PSS data presented in the database.

The shape of $g_L(r)$ for the \co Shimadzu Ralston type 1 source
(Figure~\ref{fig_new_sources}) is distinct from the other sources and
matches the expected shape for a class A source \cite{TR08,Sa20}. The dose
buildup region along the transverse axis of the source is caused by the
partial occlusion of the two radioactive pellets by the stainless steel rod
at small $r$ values. Also, the geometry factor, $G(r,\theta)$, assumes
there is radioactivity near r=0 whereas the activity is at least 4.5~mm
away.

\FloatBarrier

\begin{figure}[ht]
\begin{center}
\includegraphics[scale=0.45]{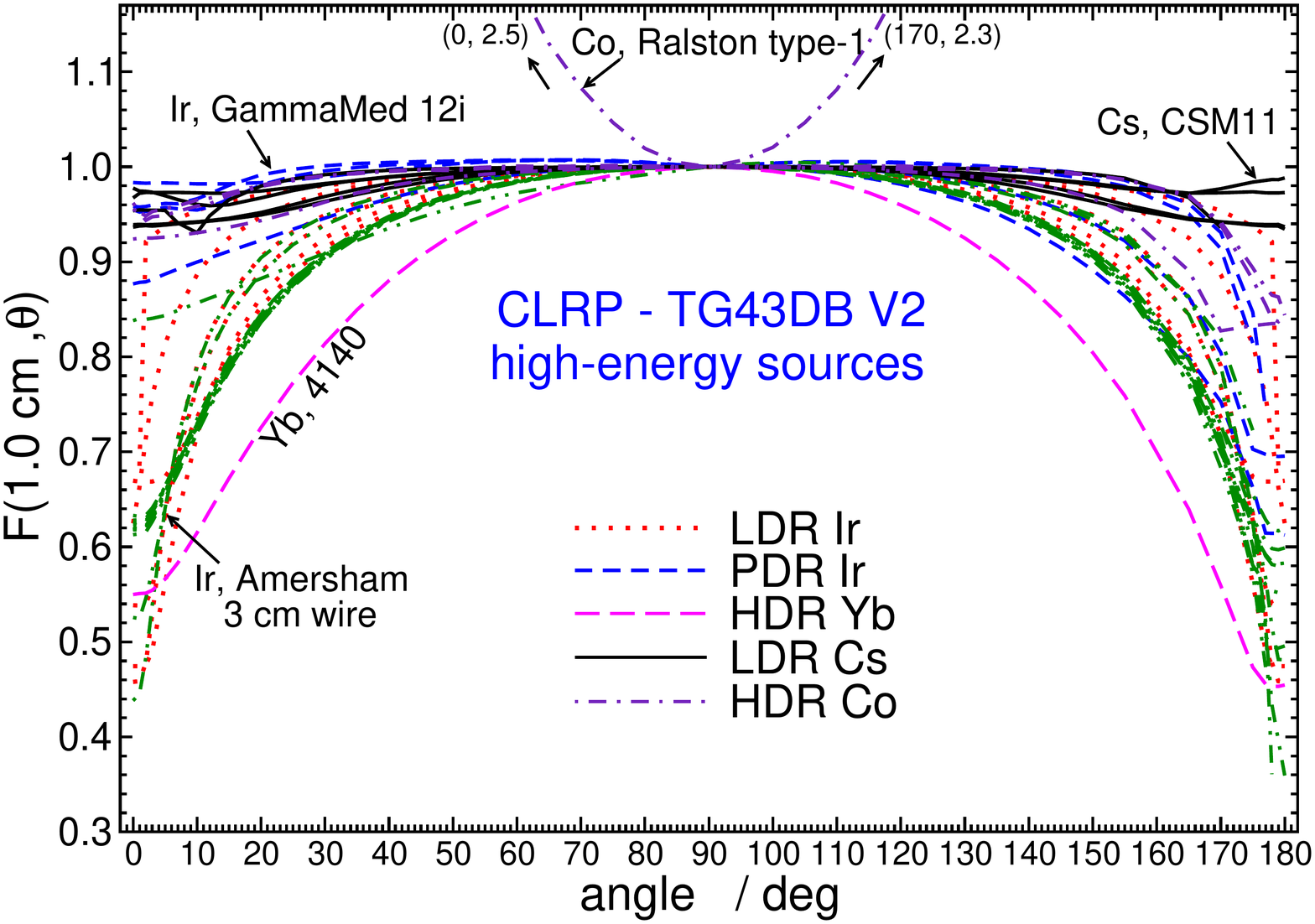}
\captionl{The 2D anisotropy function, $F(1.0~{\rm cm}, ~{\theta})$, of all 33
sources (6 LDR, 4 PDR, and 13 HDR \ir, 1 HDR \yb, 5 LDR \cec, and 4 HDR
\co) in the HE CLRPv2 database as a function of radionuclides and angle
(degrees). Electron transport just included for \co sources.
\label{fig_Frtheta_all} }
     \end{center}
   \end{figure}

In section~\ref{sec-edep-Fofrt}, the potentially dramatic effects of 
electron transport on  F($r,\theta$) values at small radii are pointed
out. Here,
Figure~\ref{fig_Frtheta_all} summarizes F(1~cm,$\theta$) data
for all sources in the \Tvv ~database. In general, in proximal
($0^{\circ}$) or distal ($180^{\circ}$) regions, anisotropy is
increased due to increased attenuation in thicker ends of many source
capsules.
The most anisotropy is for \yb sources and the least anisotropy is for the
\ce sources, followed by \coc, and then \ir sources. 

The \co (Ralston type 1) source has a distinctive anisotropy function
which, close to $0^{\circ}$ or $180^{\circ}$ is $>2.5$ for
$r=1$~cm since these points are
very close to the two active pellets of the source
(Figure~\ref{fig_new_sources}). However at $r=5$~cm it is similar to 
other sources with a value near 0.9 (see webpages).\\

\section{Data Format and Access}
\label{format}
The \Tvv~ website is hosted at Carleton University, Ottawa, Canada at
\url{https://physics.carleton.ca/clrp/egs_brachy/seed_database_HDRv2} or
\url{http://doi.org/10.22215/clrp/tg43v2}. The main webpage of the database
lists the 33 HE  and 40 LE brachytherapy sources for which the online
datasets are available as well as details about source spectra, half-lives,
average energies, voxel resolutions used in MC calculations, and a
spreadsheet of $S_K^{\rm hist}$ values as defined just above
eqn~\ref{eq_units}.  The database includes all of the same types of data as
discussed regarding the LE sources in the \Tvv~ database and enumerated in
our previous database paper\cite{Sa20}. The exception is that for the HE
sources, spectra and PSS data are included for all sources rather than just
representative sources as done for the LE database.

\newpage
\section{  Potential Impact}\label{impact} 
Currently most clinical dosimetry of HE sources is based on the TG-43
formalism.  The fully benchmarked  \Tvv~ source models developed here will
be distributed freely with the \eb distribution at
\url{https://github.com/clrp-code/egs_brachy} for research dosimetry and MC
dose calculations in treatment planning or retrospective studies.  The PSS
data in the database supports the TG-186~\cite{tg186} recommendations for
model-based dose calculations and dose heterogeneity calculations for
treatment planing.

The calculated dose-rate constants reported here can be used to support
future recommendations for clinical use. They have much better statistical
precision than many previous studies. 
This database may be updated in the future to add dosimetry datasets for
new HE sources.  In view of the fact that the HE \Tvv~ values reported here
are close to  the previous \Tv~ data, the most important impact here is
validation of the \eb source models as well as documentation.

\section{Conclusion}
The entire  LE and HE \Tvv~
database includes the datasets for 73 sources compared to 44 sources in the
\Tv~ database, and are available via \url{http://doi.org/10.22215/clrp/tg43v2}. 
The \Tvv~ data are validated in this paper by comparison of \eb dose-rate
constant data with \BD data in
\Tv~ and any other literature data which exists. 
Results are in good agreement with previous DRC results. The database also
includes extensive comparisons to previous \gofr~ and \Fofrt~ data. 
The current work improves statistical uncertainties, source volume
corrections, and some source geometry models compared to the \Tv~
calculations. It is shown that electron transport must be modelled 
for \co sources although this is not essential for lower-energy
sources except very close to the source.
The validated \eb models of all 33 LE and 44 HE
sources will be freely distributed with \eb distribution 
enabling more accurate brachytherapy dosimetry research and
advanced model-based dose calculations.

\section{Acknowledgements}
This work was supported by the Natural Sciences and Engineering Research
Council of Canada, the Canada Research Chairs program, and the Ministry of
Research and Innovation of Ontario. The authors thank Harry Allen for
assistance running early calculations and Facundo Ballester for
providing detailed source dimensions for the BEBIG HDR sources.


\section*{References}
\addcontentsline{toc}{section}{\numberline{}References}
\vspace{-1.5cm}
\setlength{\baselineskip}{0.43cm}	



\end{document}